# Homogenization of Halide Distribution and Carrier Dynamics in Alloyed Organic-Inorganic Perovskites

*Juan-Pablo Correa-Baena,[1]\* Yanqi Luo,[2] Thomas M. Brenner,[2] Jordan Snaider,[3] Shijing Sun,[1] Xueying Li,[2] Mallory A. Jensen,[1] Lea Nienhaus,[1] Sarah Wieghold,[1] Jeremy R. Poindexter,[1] Shen Wang,[2] Ying Shirley Meng,[2] Ti Wang,[3] Barry Lai,[4] Moungi G. Bawendi,[1] Libai Huang,[3] David P. Fenning,[2]\* and Tonio Buonassisi[1]\**

**Affiliations:**

[1]Massachusetts Institute of Technology, 77 Massachusetts Avenue, Cambridge, MA 02139
[2]Department of Nanoengineering University of California San Diego La Jolla, CA 92093, USA
[3]Department of Chemistry, Purdue University, West Lafayette, IN 47907, USA
[4]Advanced Photon Source, Argonne National Laboratory, 9700 Cass Avenue, Lemont, IL 60439

JPCB and YL contributed equally to this work.

\*Correspondence to: JPCB jpcorrea@mit.edu, DPF dfenning@eng.ucsd.edu and TB buonassisi@mit.edu

**Abstract**:

Perovskite solar cells have shown remarkable efficiencies beyond 22%, through organic and inorganic cation alloying. However, the role of alkali-metal cations is not well-understood. By using synchrotron-based nano-X-ray fluorescence and complementary measurements, we show that when adding RbI and/or CsI the halide distribution becomes homogenous. This homogenization translates into long-lived charge carrier decays, spatially homogenous carrier dynamics visualized by ultrafast microscopy, as well as improved photovoltaic device performance. We find that Rb and K phase-segregate in highly concentrated aggregates. Synchrotron-based X-ray-beam-induced current and electron-beam-induced current of solar cells show that Rb clusters do not contribute to the current and are recombination active. Our findings bring light to the beneficial effects of alkali metal halides in perovskites, and point at areas of weakness in the elemental composition of these complex perovskites, paving the way to improved performance in this rapidly growing family of materials for solar cell applications.

**One Sentence Summary**: Alkali metals in lead-halide perovskites are observed to agglomerate in highly concentrated aggregates, which coincide with uniform distribution of the halides I and Br, yielding homogenous electronic dynamics and improved solar cell performance.

## Introduction

Emerging solar cell technologies based on thin films and simple deposition methods promise to reduce production cost and produce high-quality semiconductors.(*1, 2*) Lead halide perovskite solar cells (PSCs) have emerged as an exciting candidate. In just a few years, PSCs have achieved power conversion efficiencies (PCEs) similar to commercial thin-film solar cells, achieving 22.7% in 2017.(*3*) Perovskites have an $ABX_3$ formula that is typically comprised of a monovalent cation, $A$ = cesium ($Cs^+$), methylammonium (MA); formamidinium (FA); a divalent metal $B$ = ($Pb^{2+}$; $Sn^{2+}$); and a halide anion $X$ = ($Cl^-$, $Br^-$; $I^-$).(*4*) The highest reported efficiencies have been achieved with perovskites with mixed MA and FA $A$-site cations, and Br and I $X$-site anions.(*5-7*) More recently, Cs has been used as the A-site cation to explore more complex compositions, including Cs and MA, Cs and FA, and Cs, MA, and FA.(*5, 8-12*) Similarly, Rb has been added into a multi-cation perovskite, showing improved efficiency(*13, 14*) and long-term device stability at elevated temperatures of 85˚C.(*13*) Rb has been suggested to increase charge carrier lifetime and mobility of the perovskite films, which help explain the improved device performance.(*15*) More recently, K was also used to improve the stabilized PCE of PSCs.(*16, 17*) These monovalent alkali metals have shown great promise to improve efficiency and stability of PSCs.

Despite these impressive results, the mechanisms that form the basis for improved electronic properties, performance, and stability, upon addition of alkali metals, are not yet well-understood. Past work hinted at the suppression of low-dimensional wide-bandgap polymorphs, which may act as recombination-active sites.(*10, 13, 15*) The structural, elemental, and electronic properties of these multi-$A$-site cation compounds, as well as solar cell performance metrics, have been investigated by bulk techniques(*8-10, 13, 16*) averaged over large areas (0.2 to 1 cm²), limiting our understanding of these complex material systems. A major flaw in this approach is that minority-phase formation and elemental agglomeration at the nanoscale is overlooked. Therefore, there is a need to use spatially-resolved techniques that allow us to image elemental distribution and its impact on electronic properties, and device performance.

In this work, we employ mapping techniques with nanoscale resolution to elucidate the elemental distribution of the alkali metals and their relationship to electronic properties and device performance. We use synchrotron-based X-ray fluorescence imaging to identify elemental distribution in multi ion perovskites. We find that in films where alkali metals are added, the halide distribution becomes homogeneous. Transient absorption spectroscopy mapping shows that perovskite films without alkali-metal iodides (and with segregated halides) suffer from more heterogeneity in the charge carrier dynamics than the those with CsI and/or RbI. At the same time, rubidium and potassium are shown to segregate into large clusters. As the alkali metal additives exceed 1%, we observe second-phase alkali-metal-rich aggregates, which induce charge-carrier recombination at Rb-rich clusters. Charge collection is hindered by these aggregates, and the formation of these alkali-rich nanoprecipitates should therefore be avoided. Our findings are crucial to the understanding of the mechanisms that make these mixed-cation perovskite materials the most efficient for solar cell applications and give insights as to how the community should continue to improve these materials.

**Result and Discussion**

*Device characterization*

We investigated the effects of alkali metal addition to the ubiquitous MA, FA, Pb, I, and Br perovskite.(*5, 6*) CsI, RbI, and KI were added to the perovskite solution, at different molar percentages ranging from 1 to 5%. Perovskite thin films were prepared with the combinations *X*-I/Br, where *X* is the alkali metal added, and I/Br refers to the $(MAPbBr_3)_{0.17}(FAPbI_3)_{0.83}$, in solution, similar to the composition previously studied by Saliba *et al.*(*13*) The most studied perovskites in this work contain 5% CsI, 5% RbI, 5% CsI and 5% RbI, and 5% KI, which are referred to in this work as Cs-I/Br, Rb-I/Br, CsRb-I/Br, and K-I/Br. Solar cells were made with these perovskites in a glass/fluorine-doped tin oxide (FTO)/compact $TiO_2$/mesoporous $TiO_2$/perovskite/spiro-OMeTAD/Au architecture. A schematic of the device cross-section is presented in Fig. 1A. A transmission electron microscopy (TEM) cross-sectional image is presented for a PSC with I/Br (Fig. 1B) perovskite (scanning electron microscopy, SEM, cross-sectional images of several compositions can be found in Fig. S1). Regardless of composition, the perovskite layers are all around 500 nm thick, as expected.(*13*) Device preparation details are given in the SI.

The solar cell performance increases with the addition of Rb and/or Cs to I/Br (Fig. S2). Representative current density-voltage (*J*-V) curves, including the backward and forward scans, for I/Br and CsRb-I/Br are shown in Fig. 1C and D, respectively. The champion device for the I/Br perovskites yields an open-circuit voltage ($V_{OC}$) of 1.09 V, a short-circuit current density ($J_{SC}$) of 22.8 mA/cm$^2$, a fill factor (FF) of 51%, and PCE of 12.6% under AM1.5 illumination. In contrast, the champion device prepared with CsRb-I/BR, yielded a $V_{OC}$ of 1.10 V, a $J_{SC}$ of 22.7 mA/cm$^2$, a FF of 76%, and a PCE of 19%. The parameters are extracted from the backward scan, and no significant difference is observed between backward and forward scans. This is in agreement with earlier findings.(*13*) While $V_{OC}$ and $J_{SC}$ in these devices remains relatively constant with perovskite composition, FF is dramatically increased with the addition of alkali-metal iodides, with an especially large discrepancy between I/Br and CsRb-I/Br samples. This effect has been previously ascribed to the reduction in concentration of the "yellow-phase" impurity of the FA-based perovskites.(*9, 10, 18*) Interestingly, the time-resolved photoluminescence decay of the perovskites slows considerably with the addition of the alkali-metal iodides (Fig. 1E). Given this slower long tail (a proxy for reduced non-radiative recombination(*19*)) for samples containing alkali-metal iodides, and the recent reports of mobility increases by RbI addition into a similar I/Br perovskite,(*15*) it is expected that $V_{OC}$ will increase. However, it is possible that the $V_{OC}$ in our devices is limited by the charge-selective contacts, and therefore improvements in the perovskite quality may be overshadowed.

*Crystallographic information*

Grazing incidence X-ray diffraction (GIXRD) was performed to confirm the crystalline phases of the perovskite thin films in this study (Fig. 1F). The bulk structures show perovskite phases in cubic symmetry. With excess $PbI_2$ in the precursor solution, our results confirmed that addition of Rb and Cs supresses the formation of the $PbI_2$ phase. The relative peak intensity of the $PbI_2$ (001) peak at 12.63° decreases upon the addition of the RbI and CsI (Fig. 1F). Other observed impurity peaks at low angles came from a small amount of the non-perovskite 1D phases of δ-$FAPbI_3$ and $CsPbI_3$. Interestingly, we did not observe the strongest $RbPbI_3$ (212) peak at 27.47° from the laboratory X-ray diffraction.

The crystal lattice of the mixed perovskites in this study is affected by both the size of the A-site cations and the X-site halides. Small peak shifts were observed on the first perovskite peak (Plane (001)), at 14.060° (I-Br), 14.054° (Cs-I/Br), 14.047° (Rb-I/Br), and 14.046° (RbCs-I/Br), respectively. Although Rb and Cs have smaller ionic radii than MA and FA, the addition of 5% RbI and/or CsI increase the I to Br ratio in the precursor solution. As the iodide perovskite becomes more prevalent in the crystal structure, we expect the lattice constant to increase since this type of perovskite has a larger lattice constant.(*20*) The addition of CsI and/or RbI in these mixed-ion perovskites lead to a slight increase in the lattice constant and unit cell volume (Fig. 1G, see SI for

Pawley refinement and Table S1). The incorporation of additional iodine sources from the alkali metals and the suppression of the PbI$_2$ formation provided an I-rich environment for perovskite phase crystallisation. This means that identifying A-site cation incorporation is difficult from only the peak positions in main perovskite XRD peaks, as it has been traditionally been studied.(*13, 15, 21*)

Peak profile analysis was performed to better understand the effect of alkali metals in the perovskite phase. The zoomed-in figures of the major diffraction peaks from the perovskite phases are presented in Fig. S3. Further analysis shows that the addition of Rb leads to an increase in the full width at half maximum (FWHM), whereas Cs results in a decrease in the peak width summarized in Fig. 1H. This effect is particularly significant on the (001) and (002) peak profile (dotted line in Fig. 1H), which is along the inorganic bonding direction of I-Pb-I, Br-Pb-Br or I-Pb-Br. The change in FWHM relates to local strain and size effects due to the incorporation of cations with different sizes in the cubic cage, as well as the positions of I and Br across the lattices to accommodate the cations while maintaining an overall cubic symmetry. Cs and Rb are too small in size to form a lead iodide cubic perovskite on their own at room temperature according to their tolerance factor.(*13, 22, 23*)

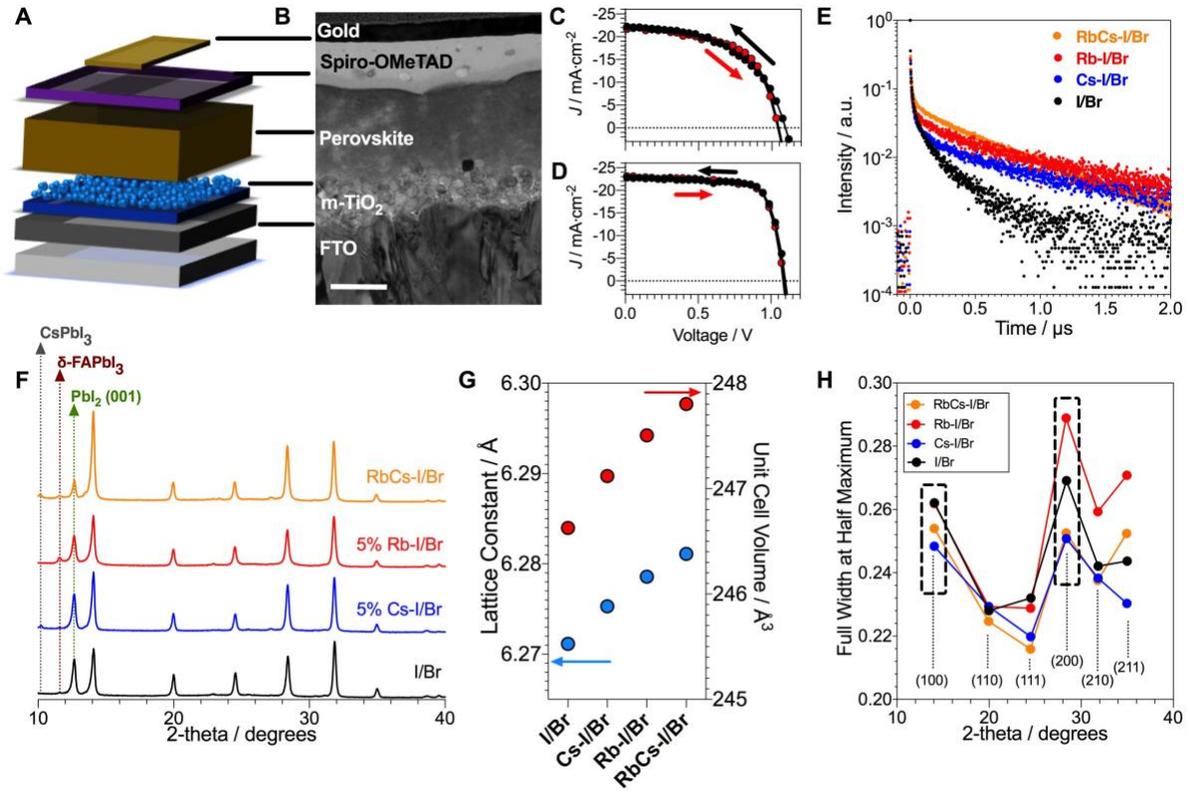

**Fig. 1. Perovskite thin film and solar cell characterization.** (**A**) Schematic of the perovskite solar cells studied, accompanied by (**B**) the cross-sectional TEM for CsRb-I/Br device; scale bar is 200 nm. Perovskite solar cell current-voltage characteristics for (**C**) I/Br and (**D**) CsRb-I/Br. (**E**) Time-resolved photoluminescence decay, (**F**) X-ray diffractograms for the studied perovskite compositions with the estimated lattice constants and unit cell volume (**G**). (**H**) Full width at half maximum as obtained from select peaks in the 10 to 40 degree range, for a the perovskite films in (**F**).

*Halide Homogenization in Rb- and Cs-Containing Films*

To understand the spatial distribution of ions of these complex multi-element perovskite thin films, we perform synchrotron-based nanoprobe X-ray fluorescence (n-XRF). The state-of-the-art materials we study are composed of mixed Br and I perovskites with small amounts of Br. Therefore, high sensitivity and resolution are needed to identify areas that are rich/poor in either halide. The resolution is limited by a 250 nm beam full-width half-maximum (FWHM) at 16.3 keV. This incident X-ray energy enables excitation of Rb $K_\alpha$, I $L_\alpha$, Br $K_\alpha$, and Pb $L_\alpha$ X-ray transitions. We first look into the distribution of the halides, as the XRD data (Fig. 1F to H) suggest major changes due to RbI and/or CsI addition. Moreover, the halide distribution in mixed halide systems has been shown to be of great importance for optoelectronic properties and perovskite device performance.(*20, 24, 25*) Hoke *et al.* showed that in mixed Br/I perovskites with high Br content, the thin films exhibit multiple emission peaks, suggesting segregation of the halides.(*24*)

Upon addition of RbI and/or CsI, we observe a homogenization in the distribution of Br, as shown in Fig. 2. The Br to Pb molar ratio XRF maps reveal the variation of bromine distribution across samples. The series of XRF maps are arranged left-to-right in order of increasing alkali ion addition. The I/Br perovskite, containing no alkali ions, shows clusters of lower Br incorporation several microns in size (Fig. 2A). Upon addition of RbI or CsI in solution, the Br distribution becomes nearly homogenous (Fig. 2B to D), particularly for the RbCs-I/Br (Fig. 2D) sample. For larger area XRF maps of the Br:Pb ratio, from which the areas shown in Fig 2A to 2D were randomly selected, see Fig. S5. Interestingly, the halide homogenization by RbI occurs for samples containing more than 1% in solution (Fig. S6). Contours around the selected Br-poor regions are drawn and shown in Fig 2A to 2D to highlight the pronounced variations. The average and the progagated quantification error of the Br:Pb molar ratios of I/Br, Cs-I/Br, Rb-I/Br, and RbCs-I/Br are $0.57 \pm 0.10$, $0.58 \pm 0.08$, $0.59 \pm 0.09$, and $0.60 \pm 0.09$ respectively, in good agreement with the expected molar ratio of the as prepared precursor solution $(MAPbBr_3)_{0.17}(FAPbI_3)_{0.83}$. The distribution of iodide in the films is found to be homogenous for different amount/type of alkali-metal iodide addition as shown in Fig. S4.

To understand the effects of halide segregation on the short-term carrier recombination dynamics of the perovskite films, we utilized transient absorption microscopy (TAM). We mapped spatial-dependent dynamics by probing the ground state bleach signal at the bandgap for I/Br, Cs-I/Br, Rb-I/Br, and RbCs-I/Br (Fig. 2 E, F, G, and H, respectively), where the relative change in TA signal (($[\Delta T(0\ ps) - \Delta T(5\ ns)]/\Delta T(0\ ps)$)) is plotted. The darker regions represent larger relative change and hence faster carrier recombination in the very short-time scale up to 6 ns. Importantly, the contrast in the TAM images reflects the spatial heterogeneity in carrier recombination dynamics. The different areas of halide aggregation may alter the carrier dynamics of the materials. Samples containing no alkali-metal iodides (Fig. 2E) exhibit the most significant spatial heterogeneity in carrier recombination dynamics, with fast decays in some areas and slower decays in others (Fig. S7E), in agreement with the results from the XRF data (Fig. 2A). In comparison, carrier dynamics in samples containing alkali-metal iodides (Figure 2 F-H) are shown to be much more spatially homogeneous, agreeing with less halide aggregation demonstrated by n-XRF. This confirms that the halide aggregation in I/Br samples can be dispersed upon the addition of alkali metals, resulting in a more homogeneous halide distribution and uniform carrier dynamics.

Upon addition of alkali-metal iodides, the TA signal loss at t = 5 ns is higher for the alkali halide-containing samples than for than for the I/Br analogues; however, all samples present this short-time quenching (for sample time-resolved TA traces see Fig. S7), in line with the photoluminescence decays presented in Fig. 1E. It is important to note that while this initial decay (first few nanoseconds) is quenched faster for the perovskites with alkali metal addition, their longer-range dynamics (TRPL in Fig. 1E) are considerably slower. This faster decay upon addition of alkali-metal iodides may be due to increased disorder in the crystal structure (as suggested by the XRD peak shifts and widening in Fig. 1F to H), which in turn have been shown to change the transient absorption dynamics of perovskites.(*26*)

We quantified the changes in Br distribution by identifying the Br-poor regions and calculating the area fraction of the Br-poor regions with respect to the overall mapped region. The area fraction is then used as the parameter to compare the relative variation of Br:Pb across sample with different chemical compositions as shown in Fig. 2I (for details see Fig. S8). The Br-poor area fraction for the I/Br perovskite is 0.14, whereas the Cs-I/Br, Rb-I/Br, and RbCs-I/Br samples exhibit area fractions of 0.007, 0.036, and 0.002, respectively. It is important to note that the homogenization of the halides is efficiently achieved as CsI is added, whereas this effect is less pronounced with the addition of RbI alone. To exemplify the variations in signal of the TA maps, we selected data points from a few lines of the maps in Fig. 2E to H, and graphed the boxplots in Fig. 2J. The I/Br sample shows a large range of TA dynamics ranging from short- to long-lived decays, whereas samples with added CsI and/or RbI show much narrower distribution, as expected from the maps in Fig. 2E to H. Quantification of the n-XRF and TA distributions provide corroborating evidence of reduced elemental and electronic heterogeneity in samples with alkali-metal ion incorporation. Solar cell results shown in Fig. S2 for all of these compositions shows that FF is the parameter most affected by the incorporation of alkali cations. Given the higher area fraction of Br-poor regions in the devices without alkali-metals added that show poorer FF, it is possible that local Br segregation can slow charge carrier extraction. This could happen due to different bandgaps forming locally and preventing charge extraction due to misaligned bands,(*27, 28*) and/or due to different mobilities for the different compositions.(*29*) Br-rich perovskite devices have been shown to exhibit multiple bandgaps and low performance metrics.(*20*)

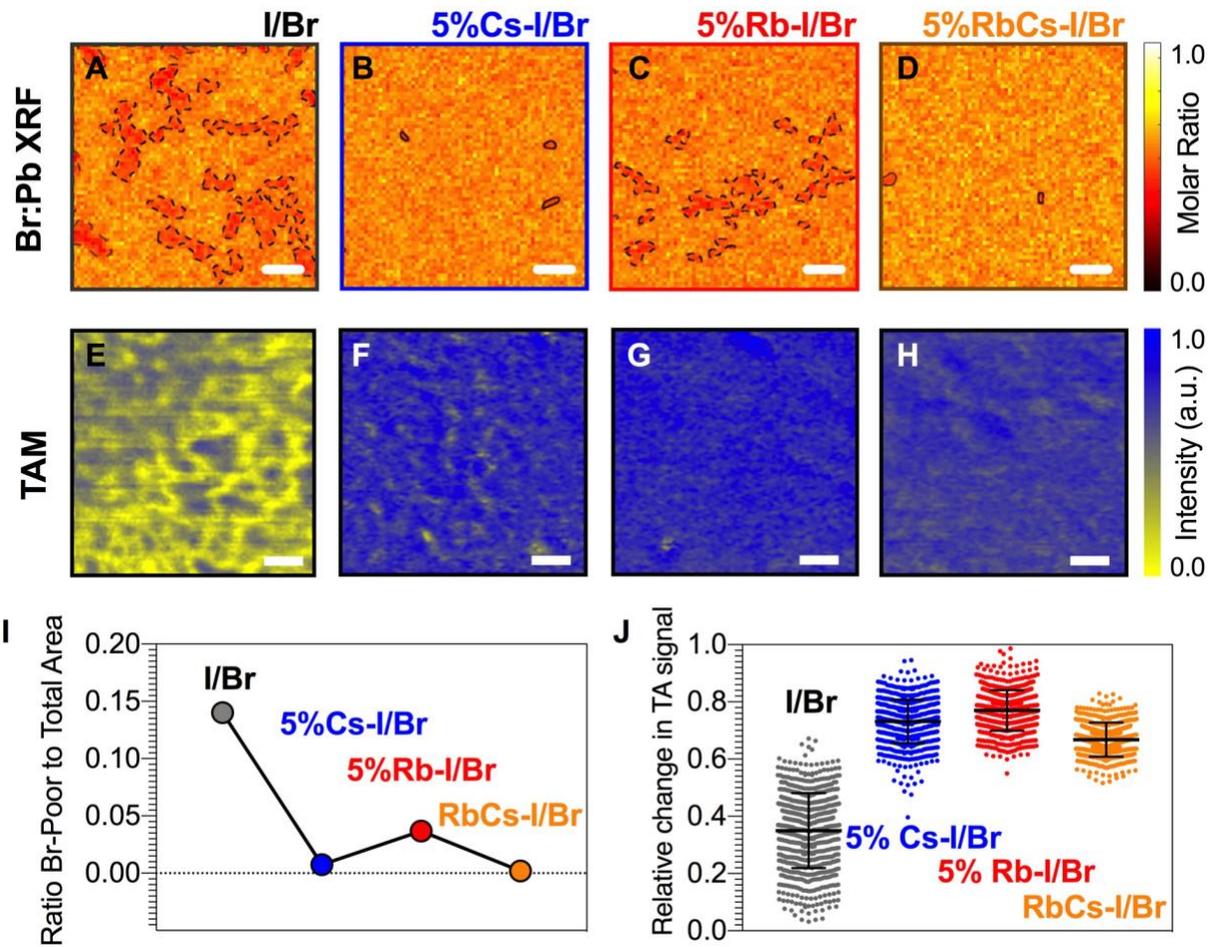

**Fig. 2. Elemental distribution and recombination dynamics.** (**A-D**) halide distribution by X-ray fluorenscence imaging, indicating homogeneous distribution of Br upon alkali metal incorporation. Br-poor areas have been marked. (**E-H**) Transient absorption microscopy (TAM) images of the relative change in TA signal [ΔT(0 ps)-ΔT(5 ns)]/ΔT(0 ps), mapped as function of sample location. The pump wavelength is at 700 nm with a pump fluence of 10 μJ/cm$^{-2}$. (**I**) Area fraction of the non-stoichiometric, Br:Pb-poor regions to the whole XRF map and (**J**) the relative change of TA signal in samples with different alkali ion additions. All scale bars are 1 μm.

*Identifying Rb and K Aggregates*

One of the most important questions regarding multi-cation perovskites is how the cations are distributed once crystalized into thin films. Several reports now show that adding Rb and Cs yields higher stabilized efficiencies,(*10, 13-15, 30*) and there is some evidence by solid-state NMR that Rb and K are not incorporated into the lattice of the perovskite structure.(*31*) We used n-XRF to map the Rb concentration and distribution within the perovskite absorbers in fully functional devices (Fig. 3A to 3C). Evaluation of the Cs elemental distribution is challenging in the iodide containing perovskites due to the peak convolution between Cs_L and I_L of the XRF emission lines. The Rb quantification details can be found in the methods section in the supporting information.

Loading of 1% Rb or greater into the perovskite absorber causes segregation of Rb-rich clusters. However, small amounts of Rb are detected in the 1%Rb-I/Br device, with a background concentration of 6 µg/cm$^2$, and some precipitation of highly concentrated Rb clusters (Fig. 3A). With a further increase in the Rb concentration, large size aggregates/dendrites appear, as seen in Fig. 3B. By introducing Cs (forming the composition CsRb-I/Br), the Rb aggregates change from large dendritic structures to compact precipitates of higher density.

Similarly, I/Br perovskite samples (deposited as thin films on FTO glass) containing varying amounts of KI showed aggregates of high concentration of K at a concentration of 5%, and needle-like structures at a concentration of 10% (Fig. S9). No considerable signal was detected for samples containing 1% KI, which we attribute to the detection limits of the low-energy K X-ray fluorescence. While some of the K may be incorporated in the film,(*16, 32*) or serve as passivant,(*17*) it is obvious from our findings that most of the K agglomerates in large clusters.

*Electronic impact of alkali-rich nanoprecipitates*

To reveal the electronic role of these large aggregates of alkali metals in the now ubiquitous Rb-I/Br samples, n-XRF/X-ray-beam-induced current (XBIC) was conducted for the devices containing different RbI concentrations. In the XBIC measurements, the incident beam generates free carriers after a cascade of thermalization interactions.(*33, 34*) These free carriers are swept out of the device as current at short circuit conditions in the functional solar cell device. Rastering of the X-ray beam allows nanoscale mapping of the collected photocurrent, where low induced current implies poor local carrier collection and vice versa. (*34, 35*) The Rb distribution and the corresponding normalized XBIC electronic profiles are presented in Fig. 3D to 3F. The XBIC profiles are normalized to the maximum current of each map, allowing cross-sample comparison. The electronic impact of the Rb particles is easily recognized in the 5% Rb loading device by comparing the Rb distribution in the XRF map and the related current collection variations in its XBIC profile, in which the poor XBIC region follows the shape of the 5-10 μm dendrite-like Rb rich aggregates (Fig. 3B and 3E). The *in-situ* measurement of n-XRF/XBIC suggests clearly that these Rb aggregates suppress charge collection in carrier dynamic pathways in the 5%Rb-I/Br device. Local XBIC deficits are seen in the devices with 1% Rb and 5% RbCs loadings, in which the Rb aggregate size is less than 1 μm, but the correlation is not as clear at the 5%Rb-I/Br device. The phenomenon of Rb hindering current collection from *in-situ* n-XRF/XBIC measurements is further verified using planar view electron-beam-induced current (EBIC) characterization.

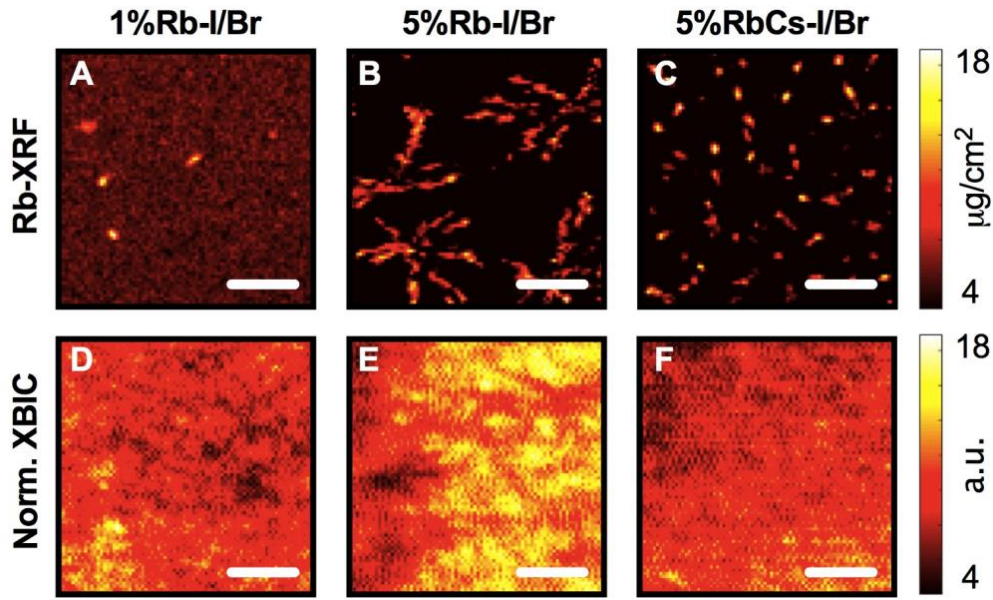

**Fig. 3. Elemental distribution of Rb and its electronic role in charge collection**. (**A** to **C**) Rb XRF maps of the alkali-added samples (1%Rb, 5%Rb, and 5%RbCs). (D to F) Normalized simultaneously collected *in-situ* X-ray-beam-induced current (XBIC) mapping while collecting the Rb XRF maps. The length of scale bar is 5 μm.

Because the XBIC signal shows low signal-to-noise ratio, we performed EBIC, where, similar to XBIC, an incident electron beam is rastered across the device generating free carriers *via* electron-matter interactions. A correlative study using EBIC and n-XRF was conducted to investigate the electronic impact of the Rb aggregates more thoroughly. The plan-view EBIC profile of the Rb-I/Br sample was collected and used to create fiducial markers by focused ion beam (FIB) for subsequent n-XRF investigation. Two regions, of lower EBIC and higher EBIC, are marked as shown in Fig. 4A. The corresponding Rb XRF map is shown in Fig. 4B, where the dark circles are the FIB markers. To assess the correlation between Rb distribution and local carrier collection, the EBIC map and the normalized Rb elemental map are divided into 10 x 10 boxes, and the average value within each box of the EBIC current is plotted against the corresponding average Rb concentration in Fig. 4C. Fig. 4C displays a negative correlation ($r^2 = 0.6$) between charge collection and Rb concentration.

It is important to consider the nature of the current suppression by Rb aggregates. Current suppression in EBIC has several possible origins: low carrier generation, current blocking, and/or carrier recombination active sites. The correlative study allows us to compare the Rb aggregate size and its current collection profile directly to evaluate whether the aggregates are recombination active. We notice that the low EBIC features in Fig. 4A are broader than the Rb ones in XRF. This broadening can come from two sources: (1) broadening due to the larger size of the carrier generation volume under the electron beam excitation relative to the n-XRF probe and/or (2) broadening due to carrier diffusion away from this generation volume, followed by recombination of free carriers near or at the Rb clusters. If the Rb aggregates are only charge-blocking and recombination *inactive*, then we would expect the EBIC maps to correspond with the Rb n-XRF maps once we account for the smearing due to a large carrier generation volume in EBIC. We modeled electron beam broadening by convoluting the Rb elemental distribution map with a simulated carrier generation profile under the electron beam (simulated using CASINO(*36*); details, beam profile, and simulated map in SI and Fig. S10A,C). Alternatively, if the Rb aggregates are recombination *active*, then their interfaces will act as a sink for diffusing excited carriers, and we would need to account for carrier diffusion to the aggregates. To model broadening due to free carrier diffusion and recombination, we further convoluted the beam-broadened Rb map with a Gaussian diffusion profile with a diffusion length of 600 nm. Upon addition of this simulated charge-carrier diffusion (Fig. 4D), the broadening of the signal approaches that of the EBIC (Fig. 4E). To illustrate the importance of accounting for carrier diffusion to the Rb aggregates further, a randomly-selected line profile is highlighted in Fig. S11A and B (blue dotted line). Modeled beam- and diffusion-broadened line profiles are plotted against the n-XRF and EBIC profiles in Fig. S11C. Comparison of the profiles indicates that while strong contrast at the n-XRF Rb features are maintained by convolution with the beam profile, this contrast is smeared out by convolution with the diffusion profile. This smearing is observed in the actual EBIC profile, indicating that carrier diffusion is critical to understanding the low-current EBIC areas corresponding to Rb-rich aggregates. Additionally, cross-section EBIC (Fig. 4F) and the corresponding SE/BSE and EDX images suggest that Rb aggregates are likely to nucleate and form near the $TiO_2$ interface and that the aggregate itself is EBIC-inactive and current-blocking. Altogether, these data suggest that Rb aggregates are optoelectronically inactive, current-blocking, and possibly recombination active, and are therefore detrimental to device performance. The entitlement in induced current available if the EBIC poor regions related to the Rb aggregates were remediated is calculated for the 1% Rb-I/Br, 5% Rb-I/Br and 5%RbCs-I/Br samples (Fig. S12). All samples show room for improved performance by addressing alkali-rich precipitates, with the largest current gains available in the samples with 5% Rb-I/Br (8% relative). This helps us understand the limitations of adding these alkali-metal halides as we aim to achieve high current collection. It should be acknowledged that this detrimental impact occurs in addition to any advantageous effect on halide homogenization the Rb clusters may have.

To understand the crystal structure of the Rb clusters, we used a synchrotron-based nano-diffraction technique to enable detection of the secondary phase locally. A parallel X-ray beam at 8 keV with 50 μm FWHM is used to collect the diffraction spectrum of this region of interest as shown in Fig. 4G. The XRD of other possible precursor phases and rubidium perovskites are shown in Fig. S13 The small peak at 27.48° is found to belong to the RbPbI$_3$ phase. Scanning X-ray Bragg diffraction microscopy (SXDM) with ~150 nm beam spot size was then used to map the low and high EBIC regions at the RbPbI$_3$ Bragg angle (2θ = 27.48˚) to understand the spatial distribution of this phase (Fig. 4H and 4I). With this nano-diffraction setup, the (212) RbPbI$_3$ crystal planes must be oriented out of plane to be detectable. Although only part of the Rb dendrites meet this criterion and diffracts in Area 1, the SXDM maps clearly exhibit the presence of the secondary phase RbPbI$_3$ in the Rb-rich aggregates, while no RbPbI$_3$ diffraction is detected in the high EBIC Area 2.

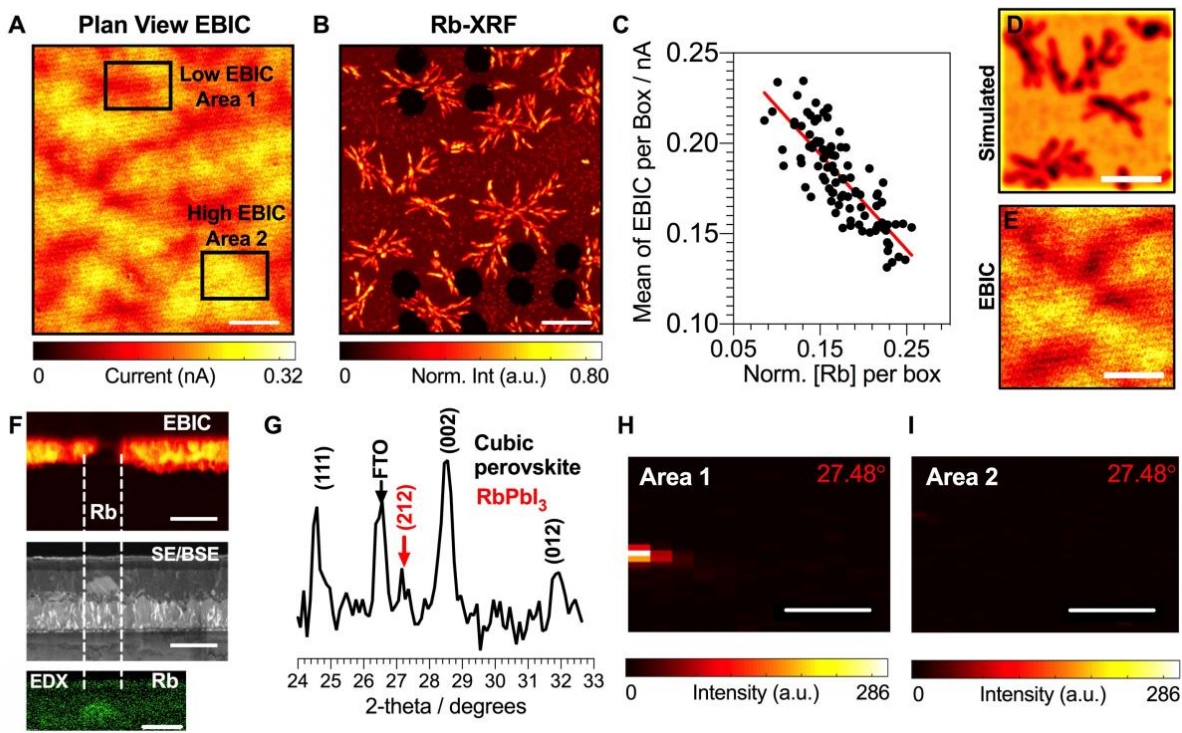

**Fig 4. The chemical nature and the extended electronic property of Rb aggregates.** (**A**) Plan view EBIC profile and (**B**) the correlated Rb-XRF map of 5%Rb-I/Br perovskite device. Three regions are marked with focused ion beam (FIB). Area 1 and Area 2 is the low and high EBIC region respectively. Length of the scale bar is 10 μm. (**C**) Scatter plot between Rb concentration and its effect on EBIC charge collection. The individual data points are produced by taking the average within 10×10 square pixel boxes in (B). (**D**) The convoluted Rb XRF map with the beam profile and Gaussian point-spread function due to diffusion with diffusion length of 600 nm. (**E**) The corresponding EBIC profile of the selected Rb XRF maps. The length of scale bar is 5 μm. (**F**) The cross-section EBIC and the simultaneously collected SE/BSE. (**G**) An X-ray diffraction spectrum taken using 8 keV beam energy with 50 μm spot size in the area within the whole region shown in (A). (**H** and **I**) The scanning X-ray Bragg diffraction microscopy map for Area 1 and Area 2 taken at the $RbPbI_3$ Bragg condition where $2\theta = 27.48°$.

**Acknowledgments:**
JPCB and YL contributed equally. JPCB conceived and designed the overall project. JPCB conducted SEM, and XRD experiments, prepared the perovskite devices. JPCB, MJ, JRP, BL, and SW conducted the initial synchrotron measurement (n-XRF). DPF, XL, and YL conducted follow-up n-XRF, XBIC, and SXDM experiments. TMB, SW, YL, and YSM conducted the EBIC, and TEM measurements. JS, TW, and LH conducted confocal transient absorption microscopy for mapping. LN and MB conducted the PL measurements. TB, DPF, and JPCB directed and supervised the research. JPCB and YL wrote the first draft of the paper. All authors contributed to the discussion and writing of the paper. The authors thank Martin V. Holt and Zhonghou Cai for helpful discussions and support in SXDM operation at APS/CNM 26-ID-C, and Pritesh Parikh for FIB sample preparation.

JPCB acknowledges support of a Department of Energy (DOE) EERE Postdoctoral Research Award. JPCB and TB are supported by the NSF grant CBET-1605495 and Skoltech 1913/R. LN was supported as part of the Center for Excitonics, an Energy Frontier Research Center funded by the US Department of Energy, Office of Science, Office of Basic Energy Sciences under Award Number DE-SC0001088 (MIT). YL, SW, YSM, and DPF are grateful for the financial support of a California Energy Commission Advance Breakthrough award (EPC-16-050). XL and DPF acknowledge the support of a Hellman Fellowship. Use of the Center for Nanoscale Materials and the Advanced Photon Source, both Office of Science user facilities, was supported by the U.S. Department of Energy, Office of Science, Office of Basic Energy Sciences, under Contract No. DE-AC02-06CH11357. The TEM work was performed at the University of California, Irvine Materials Research Institute (IMRI) using instrumentation funded in part by the National Science Foundation Major Research Instrumentation Program under Grant CHE-1338173. The FIB was performed at the San Diego Nanotechnology Infrastructure (SDNI), a member of the National Nanotechnology Coordinated Infrastructure, which is supported by the National Science Foundation (Grant ECCS-1542148).

# Supplementary Materials for

## Homogenization of Halide Distribution and Carrier Dynamics in Alloyed Organic-Inorganic Perovskites

*Juan-Pablo Correa-Baena,[1]\* Yanqi Luo,[2] Thomas M. Brenner,[2] Jordan Snaider,[3] Shijing Sun,[1] Xueying Li,[2] Mallory A. Jensen,[1] Lea Nienhaus,[1] Sarah Wieghold,[1] Jeremy R. Poindexter,[1] Shen Wang,[2] Ying Shirley Meng,[2] Ti Wang,[3] Barry Lai,[4] Moungi G. Bawendi,[1] Libai Huang,[3] David P. Fenning,[2]\* and Tonio Buonassisi[1]\**

Correspondence to: JPCB jpcorrea@mit.edu, DPF dfenning@eng.ucsd.edu and TB buonassisi@mit.edu

**This PDF file includes:**

Materials and Methods
Figs. S1 to S13
Table S1

**Materials and Methods**

### Device Fabrication

The patterned-F-doped SnO$_2$ (FTO, Pilkington, TEC8) substrates were cleaned by sonicating sequentially in 2% Hellmanex detergent in water, ethyl alcohol and acetone. A dense electron selective layer of TiO$_2$ (bl-TiO$_2$, ~50 nm in thickness) was deposited onto a cleaned substrate by spray pyrolysis, using a 20 mM titanium diisopropoxide bis(acetylacetonate) solution (Aldrich) at 450°C. A mesoporous TiO$_2$ (meso-TiO$_2$, Dyesol particle size: about 30 nm, crystalline phase: anatase) film was spin-coated onto the bl-TiO$_2$/FTO substrate using a diluted TiO$_2$ paste (5:3 paste:ethanol), followed by calcining at 500°C for 1 h in air to remove organic components.

The perovskite films were deposited from a precursor solution containing FAI (1 M, Dyesol), PbI$_2$ (1.1 M, TCI Chemicals), MABr (0.2 M, Dyesol) and PbBr$_2$ (0.22 M, TCI Chemicals) in anhydrous DMF:DMSO 9:1 (v:v, Acros). The alkali-metal iodides were added as CsI, RbI, and KI, in molar ratios of the alkali-metal to lead, ranging from 1% to 10%, as these are widely reported concentraitons in the literature. The perovskite solution was spin-coated in a two-step program; first at 1000 for 10 s and then at 6000 rpm for 30 s. During the second step, 200 µL of chlorobenzene were poured on the spinning substrate 15 s prior the end of the program. The substrates were then annealed at 100 °C for 1 h in a nitrogen filled glove box.

The spiro-OMeTAD (Merck) solution (70 mM in chlorobenzene) was spun at 4000 rpm for 20 s. The spiro-OMeTAD was doped at a molar ratio of 0.5, 0.03 and 3.3 with bis(trifluoromethylsulfonyl)imide lithium salt (Li-TFSI, Sigma Aldrich), tris(2-(1H-pyrazol-1-yl)-4-tert-butylpyridine)- cobalt(III) tris(bis(trifluoromethylsulfonyl)imide) (FK209, Dyenamo) and 4-tert-Butylpyridine (TBP, Sigma Aldrich), respectively. Finally, a 100 nm Au top electrode was deposited by thermal evaporation. The active area of this electrode was fixed at 0.16 cm$^2$.

### General Characterization

Grazing incidence X-ray diffraction was performed using a Rigaku SmartLab Diffractometer. Pawli refinements of all compounds were performed using Highscore Plus version 4. The cross-sectional images were investigated by a Zeiss Merlin field-emission scanning electron microscopy (FESEM, Zeiss). Time-resolved photoluminescence measurements were obtained by time-correlated single photon counting (TCSPC). The sample was excited by a pulsed 405 nm wavelength laser (PicoQuant LDH-P-FA-530-B) and the laser power was adjusted by a neutral density filter wheel to obtain a <5% count rate in each measurement to avoid pile-up artifacts in the detector. The repetition rate was set to 200 kHz for Rb, K and 100 kHz for Cs. The emission was collected by parabolic mirrors and focused onto a silicon single-photon avalanche photodiode (Micro Photon Devices $PD-100-C0C) and photon arrival times were recorded using a PicoHarp 300 (PicoQuant) using an integration time of 300 s. Excess laser scatter was removed by a 532 notch filter (Chroma Technology Corp.) and a 550 nm long-pass filter (ThorLabs).

### n-XRF Characterization

To obtain elemental distribution profile in perovskite absorber, functional hybrid perovskite solar cells were investigated by means of synchrotron-based nanoprobe X-ray fluorescence (n-XRF) with a 250 nm full-width half-maximum (FWHM) focused beam at 16.3 keV at beamline 2-IDD in a helium environment of the Argonne National Laboratory (APS). The *n*-XRF measurement was conducted with backside Au contact facing the incident X-ray beam with point-by-point fluorescence spectrum collected during mapping. MAPS software was used to fit and deconvolute overlapping peaks and the background from the fluorescence data.[1] The stoichiometry of Br:Pb in absorbers was then quantified using a methylammonium lead bromide single crystal.

### Cluster Analysis on the Br:Pb ratio XRF maps

To have a quantitative evaluate on the heterogenous Br distribution, a cluster analysis was exploited to identify the Br rich/poor area within the scanned window using the built-in *bwconncomp* and *regionprop* functions from MATLAB. Binary Br:Pb XRF maps are produced by applying a threshold limit of using the maps, in which, Br:Pb above and below 0.52 is considered as Br-rich and Br-poor regions, respectively. The pixel value of the Br-poor region is assigned to 1, while the pixel of the Br-rich region is assigned to 0. The binary maps of the Br:Pb ratio maps are displayed in Row 2 of Fig S8. The threshold limit was determined based of the prepared precursor mixture, yielding absorber chemistry to be $(MAPbBr_3)_{0.17}(FAPbI_3)_{0.83}$ where 0.51 is the expected stoichiometry Br:Pb ratio. The *bwconncomp* algorithm in MATLAB identifies the Br-poor regions and highlights the regions where have more than 3 "white" pixels being recognized in the red bounding boxes as shown in Row 3 of Fig S8. The *regionprop* function was further used to extract the statistic within individual bounding box. The area fraction of the non-stoichiometric Br ratio within the whole map of different alkali added samples is shown in the scatter plot of Fig 2I. The area fraction calculation is conducted by dividing the total number of the Br:Pb poor pixels to the total number of pixels of the entire map.

### Transient absorption imaging:

The transient absorption images and dynamics were acquired by a home-built, femtosecond transient absorption spectroscopy setup. The output of a high repetition rate amplifier (PHAROS Light Conversion Ltd., FWHM = 200 fs, 400 kHz repetition rate, pulse energy of 100 μJ, 1030 nm fundamental) was used to feed two independent optical parametric amplifiers (OPA, TOPAS-Twins, Light Conversion Ltd) to generate the pump beam (700 nm) and probe beam (755 nm). An acousto-optic modulator (Gooch and Housego, R23080-1) was used to modulate the pump beam at 100 kHz. A mechanical translation stage (Thorlabs, DDS600-E) was used to delay the probe with respect to the pump. Both the pump and probe beams were focused onto the sample by an oil immersion objective (CFI Apo TIRF, Nikon Inc., 60×, NA 1.49). With the pump beam being filtered out, the probe beam was collected by another objective (Nikon, S Plan Fluor ELWD 20×, NA 0.45) and was detected by an avalanche photodiode (APD; Hamamatsu, C5331-04). The change in the probe transmission (ΔT) induced by the pump was detected by a lock-in amplifier. A pair of Galvanometer mirror (Thorlabs GVS012) was used to scan the probe beam and pump beam together in space to obtain the transient signal mapping. To construct Figures S2 E-H, we plot the difference between transient bleach signal at time zero (0 ns time delay, $\Delta T_0$) and that after a 5 ns time delay ($\Delta T_5$) is compared to the signal arising from the initially generated carriers at time zero.

$$\frac{\Delta T_0 - \Delta T_5}{\Delta T_0}$$

### In-situ n-XRF/XBIC Characterization

The *in-situ* n-XRF and XBIC (X-ray beam induced current) characterization was conducted using similar setup as previous studies,[2] where X-ray beam penetrating through the backside Au metal contact allowed collection of both elemental distribution (XRF) and charge collection profile (XBIC) simultaneously with 250 nm step size and a dwell time per point of 50 ms. A lock-in current pre-amplifier at the beamline was used to collect the induced current signal for XBIC. To improve the chemistry detection further, a second "chemistry" map was collected using step size as small as 150 nm with the same amount of dwell time. The Rb incorporation was quantified using $RbI_2$ Micromatter standard with known loading concentration of 6.2 µg/cm$^2$.

### EBIC Characterization

The electron beam induced current (EBIC) was collected using an FEI Scios Dual Beam microscope with a Mighty EBIC 2.0 controller (Ephemeron Labs) and a Femto DLPCA-200 pre-amplifier.

*Planar Configuration:* Similar to XBIC measurements, the e-beam was rastered across the sample through the backside Au metal contact and the generated carriers were collected via the front/back contacts, Au and FTO. The accelerating voltage and current were 10 kV and 13 pA, respectively. The 10 kV accelerating voltage was chosen in order to allow electrons to penetrate the top contact and reach the perovskite layer. A beam current of 13 pA ensured that damage was minimized and injection was kept low. The pixel size varied between 21 nm and 84 nm depending on the magnification used, and the dwell time was 26.5 µs. EBIC and SEM images of the same region of interest were collected simultaneously as shown in **Figure S12**. The EBIC maps are normalized to the maximum current in each individual map, allowing cross-sample comparison for samples with different Rb loading. The comparison between EBIC and SEM further indicates that the variation shown in EBIC maps is not due to surface morphology nor pinholes. We observed an activation effect that depended on total electron dose. The total EBIC signal increased over several scans, before decreasing due to damage. However, the features observed in our images were not dose-dependent.

*Cross-Section Configuration:* For cross-section EBIC measurements, the sample was scribed by a diamond-wheel type glass cutter on the glass side. The sample was then cleaved by applying pressure on the device-side above the cleave with a diamond pen. The mechanically cleaved samples were then immediately loaded into the EBIC system. Imaging was performed in FEI Optiplan mode using the T2 (in-lens like) detector, at 3kV and 1.6pA. The pixel size and dwell time were 5.6 nm and 26.5 µs, respectively. We did not observe the same activation effect as in planar measurements, possible due to increased energy deposition under these conditions. We would like to stress that interpretation of cross-section EBIC measurements with mechanically cleaved devices must be done with care. The mechanical break leaves morphological artifacts that must be ruled out in the analysis. In order to obtain the reported result in Figure 3H in the main text, we established that the effect was repeatable and reliable regardless of the cross-section structure at the Rb cluster.

### CASINO Electron Beam Profile Modeling

We modeled the profile of energy deposited by the electron beam by Monte-Carlo simulation using CASINO software. The energy deposited is directly proportional to the number of free carriers generated by the electron beam. We constructed a model consisting of a gold top layer, carbon middle layer to simulate the HTL, and perovskite bottom layer. A 10 keV beam impinged perpendicular to the layer structure. A matrix of gold and HTL thicknesses were tested to assess variability in the profile due to sample variation. As the Au and HTL layers generate negligible EBIC current (as verified from cross-section measurements), we considered only the energy deposited in the perovskite layer. As EBIC images are a 2D projection of a 3D process, we projected the 3D beam profile in the perovskite layer onto the plane parallel to the layer. This produced the beam profile shown in Figure S10(a). Within the range of Au thicknesses of 80±10 nm and HTL thicknesses of 200±20 nm, this beam profile has a full-width-half-maximum of ~400 nm, without much variation with contact thickness.

### **Transmission Electron Microscopy and Focused Ion Beam Sample Preparation**
Transmission electron microscopy (TEM) images were taken with a FEI 200kV Sphera Microscope. Samples for TEM were prepared by focused ion beam (FEI Scios DualBeam FIB/SEM). The FIB-TEM sample preparation follows a previously reported procedure.[4]

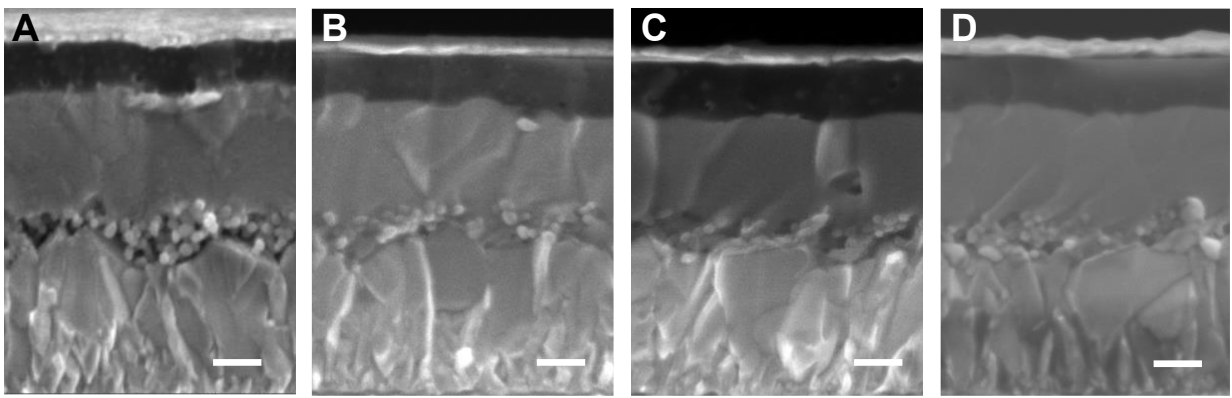

**Fig S1.** Cross-sectional SEM images of PSCs with an architecture FTO/compact TiO2/mesoporous TiO2/perovskite/Spiro-OMeTAD/Gold. The perovskite photoabsorber is varied as (A) I/Br, (B) Rb-I/Br, (C) Cs-I/Br, and (D) CsRb-I/Br. The scale bars are 200 nm.

**Fig S2.** Performance metrics of the compositions studied

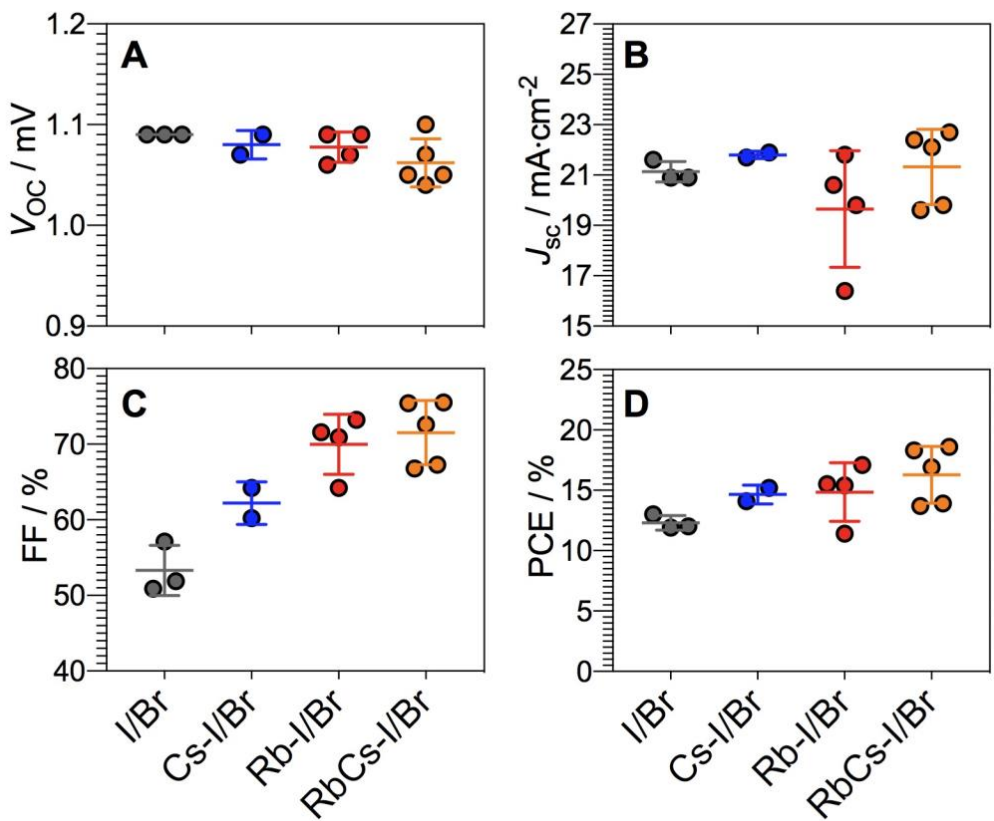

**Table S1.** Pawley refinement of lattice parameters from laboratory XRD in Fig. 1.

|           | Lattice / Å         | Volume / Å³        |
|-----------|---------------------|--------------------|
| I/Br      | 6.27114 ± 0.0002    | 246.63 ± 0.0237    |
| Cs-I/Br   | 6.27531± 0.0002     | 247.12 ± 0.0239    |
| Rb-I/Br   | 6.27859 ± 0.00025   | 247.51 ± 0.0292    |
| RbCs-I/Br | 6.28110 ± 0.0002    | 247.80 ± 0.0241    |

Out-of-plane grazing incidence X-ray diffraction was performed using a Rigaku SmartLab Diffractometer for 2-theta angles of 10° - 70° with a step size of 0.03°. Profile analysis was carried out using Topas Academic V6. The perovskite phase in the CsRb-I/Br thin-film was refined as a single cubic phase and lattice parameters were determined based on a symmetry of $Pm\bar{3}m$ following the work by Saliba *et. al*(Reference [3]). Pawley refinement (listmode) is shown in the following figure:

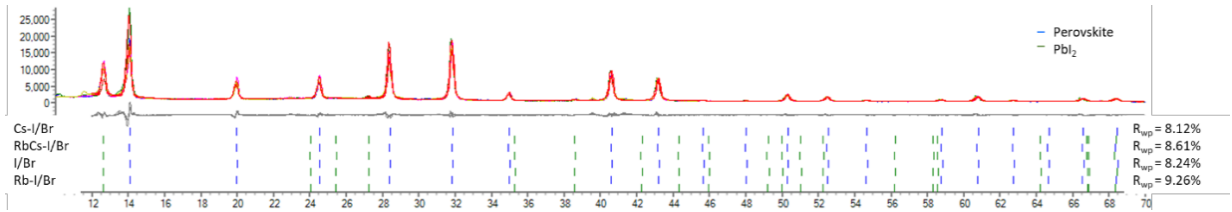

**Fig S3.** Close-ups of major laboratory X-ray diffraction peaks of mixed cation and halide perovskite phases for the thin films studied in this work.

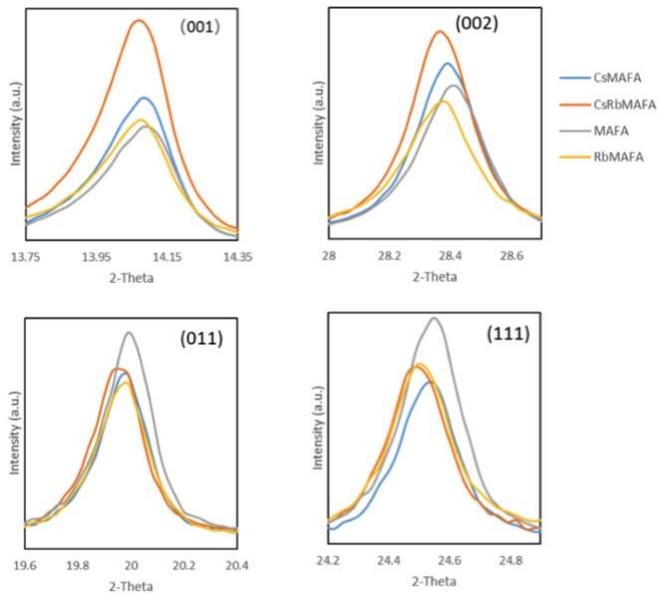

**Fig S4. Iodine elemental distribution in the samples with different alkali ions additions.** The length of scale bar is 10 μm.

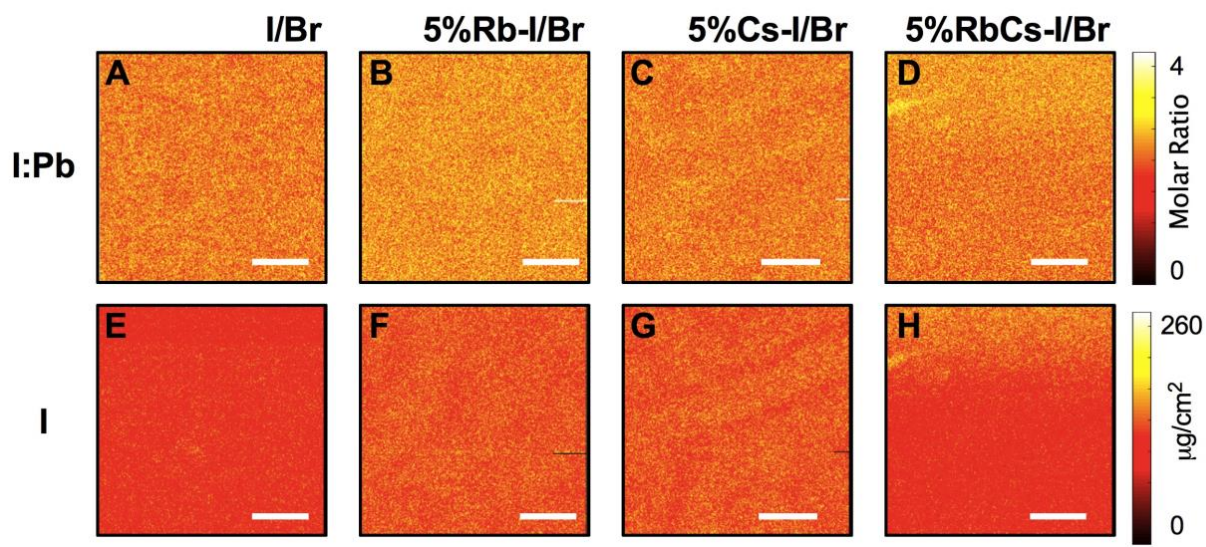

**Fig S5. Larger scale Br:Pb XRF maps for devices having different alkali ions additions.** The square boxes in (A) to (D) indicate the smaller regions shown in Fig 2 main text. Scale bars are 10 μm.

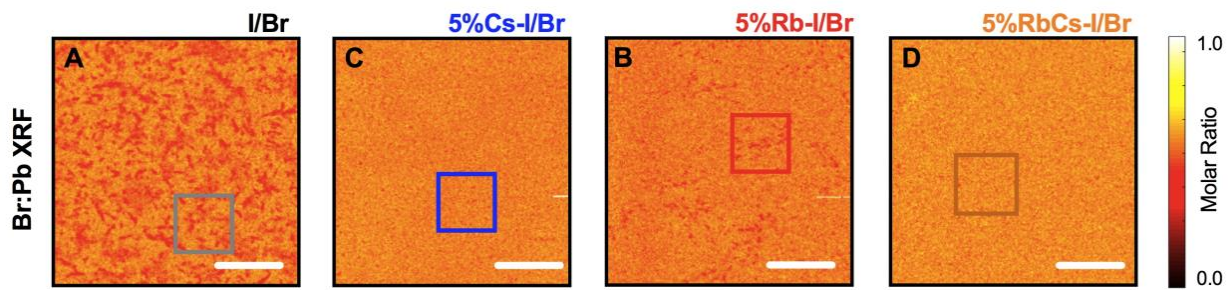

**Figure S6.** Halide homogenization as alkali ion loading increases. The Br:Pb XRF maps are used to reveal the halide homogenization in various alkali ion loadings. The alkali ion assisted homogenization effect occurs with Rb concentrations above 1% Rb is introduced. Scale bars are 10 μm.

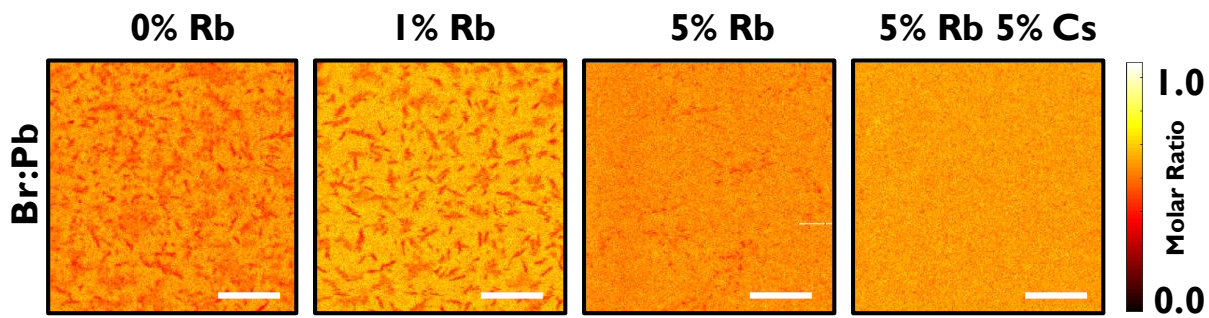

**Figure S7.** Charge carrier dynamics as measured by transient absorption microscopy (TAM). (**A-D**), the relative change in TA signal [ΔT(0 ps)-ΔT(5 ns)]/ΔT(0 ps) is mapped as function of sample location. The pump wavelength is at 700 nm with a pump fluence of 10 µJ/cm$^{-2}$. The corresponding time resolved traces below each TAM panel show carrier dynamics taken at positions as indicated. The perovskites studied are: MAFA (**A, E**), 5% CsMAFA (**B, F**), 5% RbMAFA (**C, G**) and 5% Cs 5% RbMAFA (**D, H**). All scale bars are 1 µm.

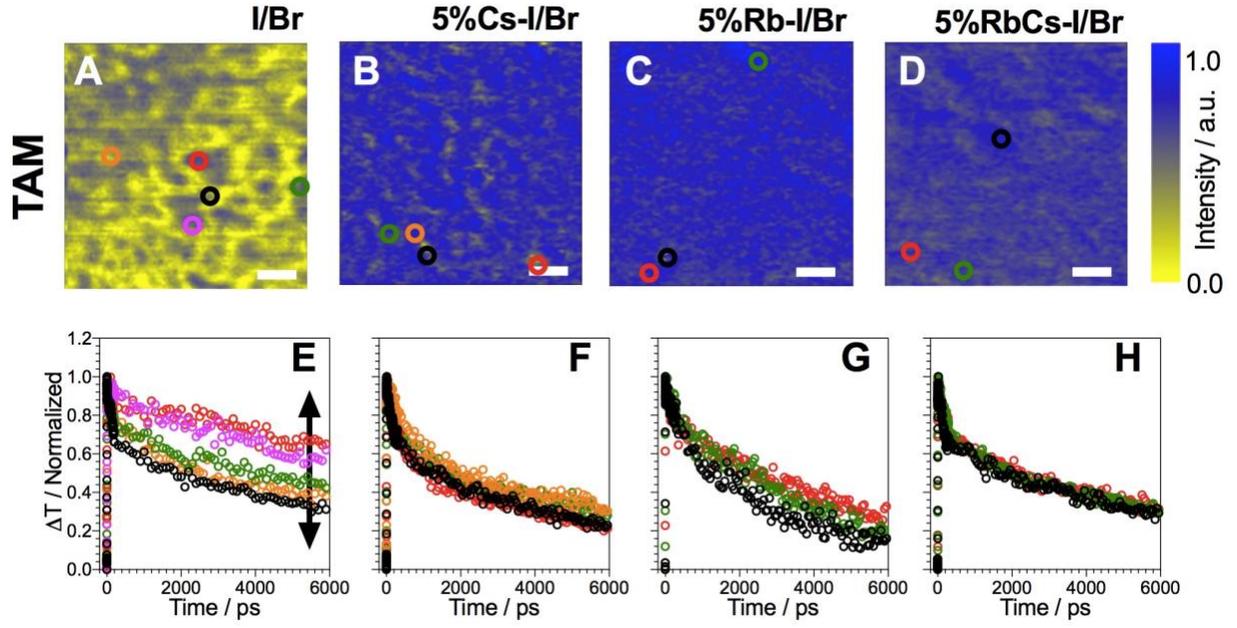

**Fig S8. Halide cluster selection in samples with different alkali-ions addition.** The Br:Pb XRF ratio maps shown in Row 1 is the same data presented in the main text in Figure 2. The Br:Pb binary maps are produced by thresholding the maps in Row 1, where pixels with Br:Pb value smaller than 0.52 is assigned to have value of 1 and values bigger than 0.52 is assigned to have value of 0 as shown in Row 2. Row 3 displays the binary maps with Br-poor regions highlighted in red bounding boxes. The length of scale bar is 10 μm. The cluster analysis for distinguishing Br:Pb rich/poor areas was carried out using *bwconncomp* and *regionprops* algorithm in MATLAB.

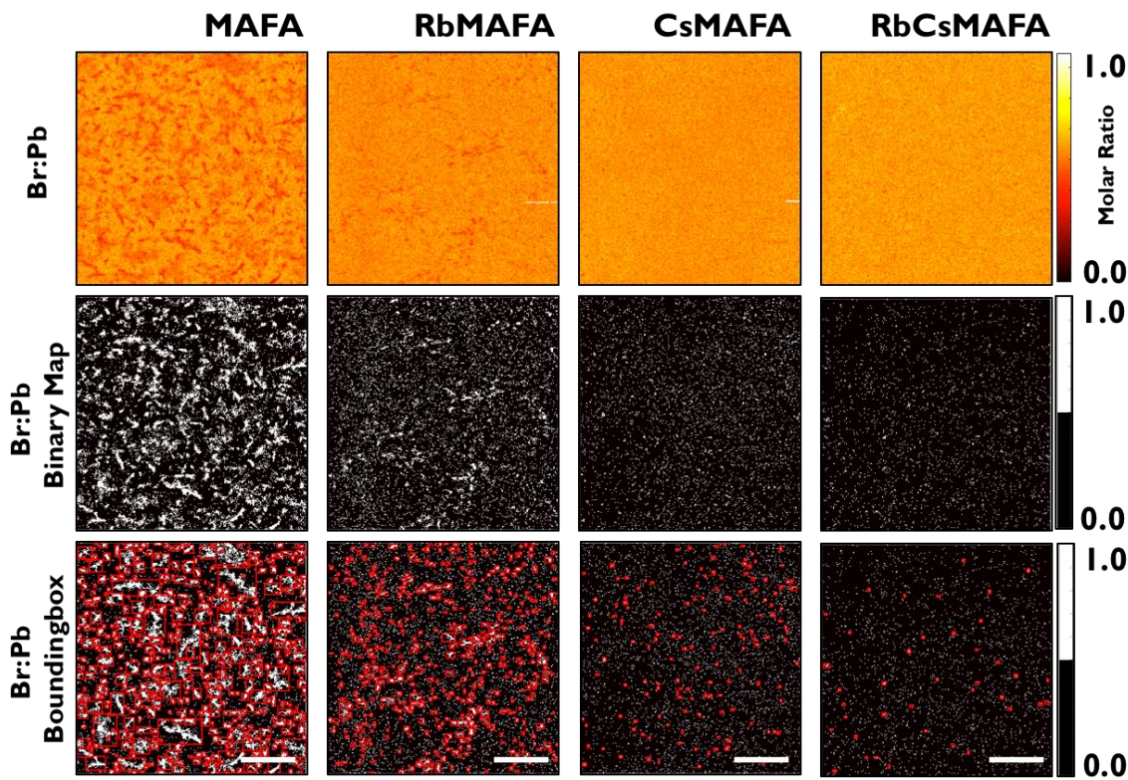

**Fig S9. Elemental distribution of Rb and K**. (**A** to **C**) Rb XRF maps of the RbI-added samples with 1%Rb, 5%Rb, and 10%Rb to I/Br perovskite. (**D** to **F**) Correlative electron beam induced current (EBIC) mapping of the samples to reveal heterogenous current collection in these samples. These samples were deposited on FTO glass substrates. The length of scale bar is 5 μm.

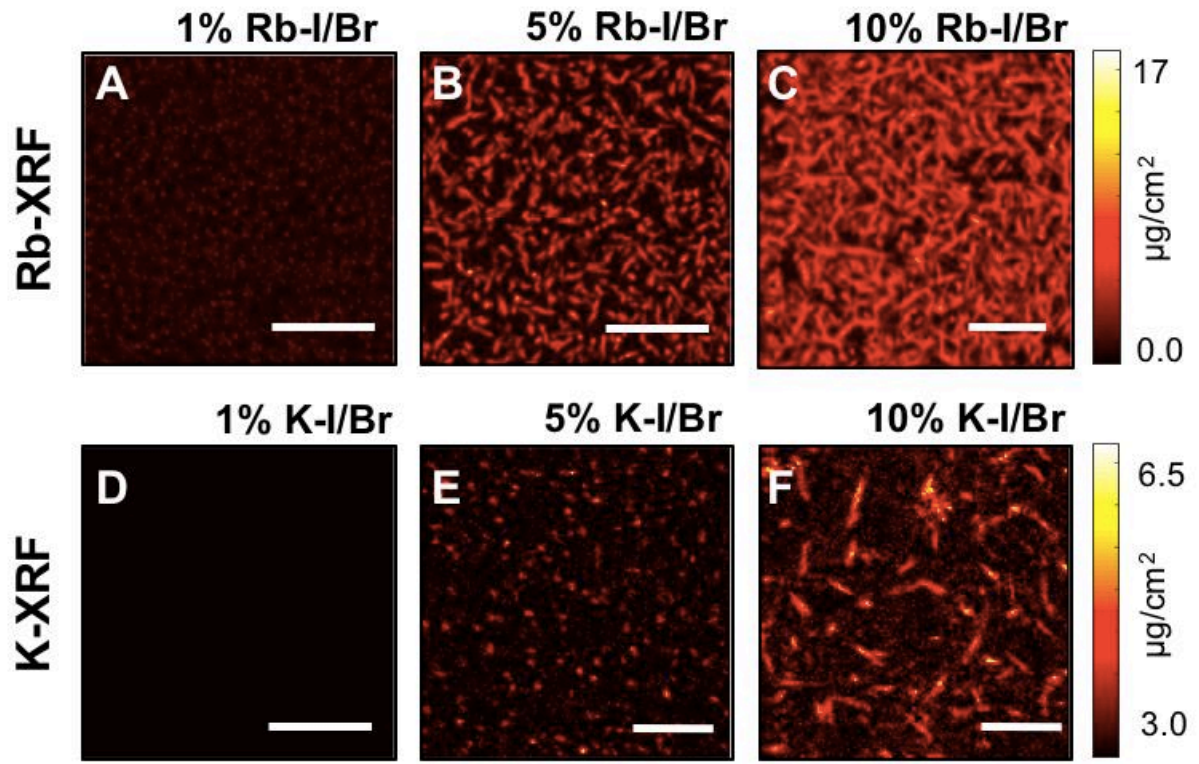

**Fig S10. The effect of electron beam and diffusion broadening on Rb clusters.** (A) Simulated profile of energy deposited in the perovskite layer by a 10 keV beam penetrating the Au and HTL layers of the device. The length of scale bar is 500 nm. (B) The Rb XRF map of the 5%Rb-I/Br sample. (C and D) The profiles of the generation volume convoluted Rb XRF maps with the simulated electron beam profile, in which D is shown in reverse color scale. (E and F) The profiles of the generation volume and gaussian diffusion convoluted Rb XRF maps with diffusion length of 600 nm, in which F is shown in reverse color scale. (G) The corresponding EBIC profile of the selected Rb XRF maps. The length of scale bar in B to G is 5 μm.

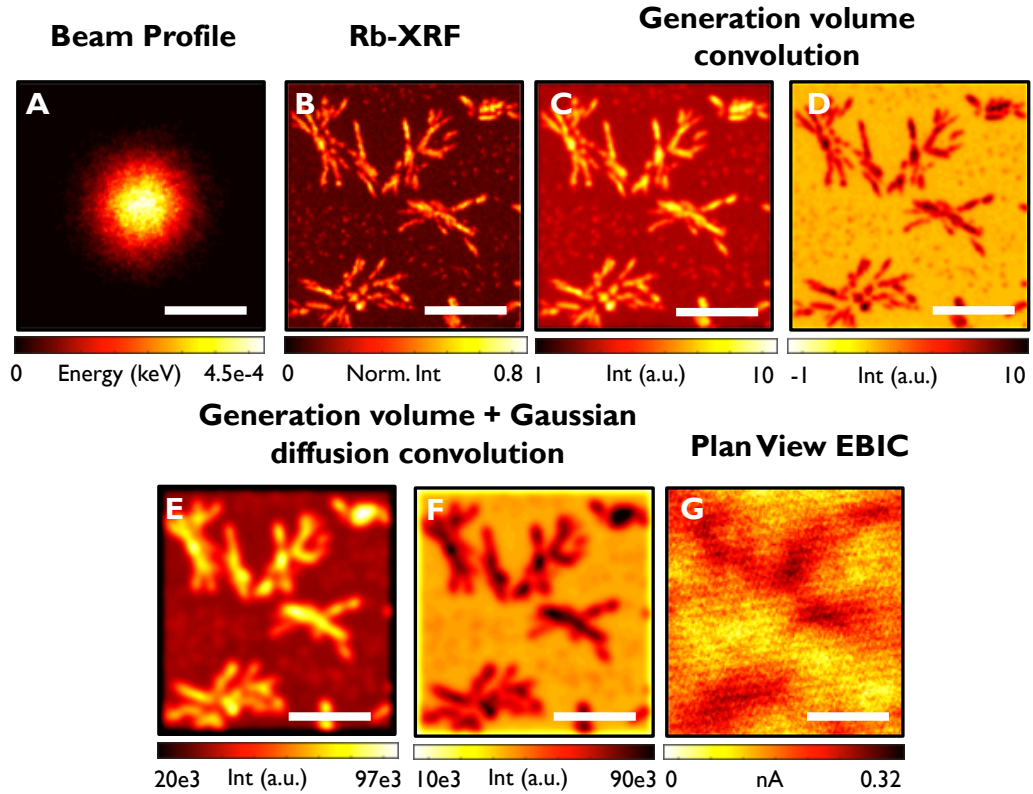

**Fig. S11. Line scan of simulated charge carrier diffusion and experimental EBIC.** (**A**) Plan view EBIC profile and (**B**) the correlated Rb-XRF map of 5%Rb-I/Br perovskite device. Three regions are marked with focused ion beam (FIB). Area 1 and Area 2 is the low and high EBIC region respectively. Length of the scale bar is 10 μm. (**C**) Displays four different line profiles. The XRF and EBIC profiles are obtained from (A) and (B). The other two profiles, XRF+Beam and XRF+Beam+Diffusion, are simulated by convoluting the XRF profile with Ebeam FWHM 400 nm and 600 nm diffusion length.

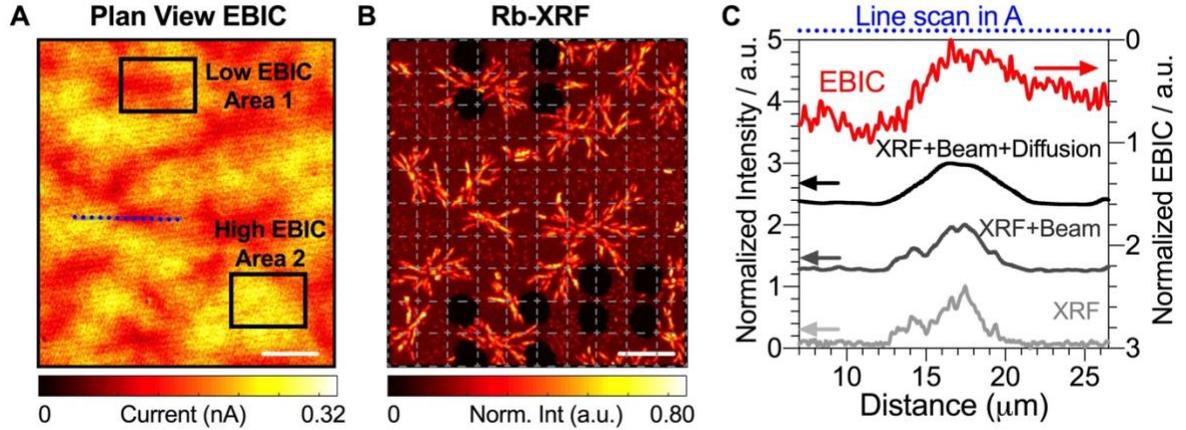

**Fig. S12. Quantifying the induced current loss from EBIC poor regions related to the Rb aggregates.** The EBIC current is adversely affected by Rb aggregates as seen in EBIC maps of the (A) 1%Rb-I/Br, (B) 5%Rb-I/Br, and (C) 5% RbCs-I/Br samples. Corresponding SEM images in for each sample are shown in (D-F). SEM reveals no pinholes or systematic variations in morphology. The EBIC is severely compromised in 5% Rb-I/Br samples, whereas the 1% RbI and RbCs-I/Br samples show less of an impact of the Rb precipitates on current collection. The maximum current entitlement in relative % that could be captured if the negative effects of the Rb precipitates were removed completely is shown in (G) and calculated according to:

$$Max\ EBIC\ Entitlement = \frac{\bar{i}_{bulk} - \bar{i}_{poor}}{\bar{i}_{bulk}} f_{poor}$$

Where $\bar{i}_{bulk}$ is the average EBIC in the bulk of the film away from the Rb aggregates, $\bar{i}_{poor}$ is the average EBIC in the poor-current collecting areas affected by the Rb aggregates, and $f_{poor}$ is the area fraction of the film affected by the Rb aggregates. The scale bars are 10 µm.

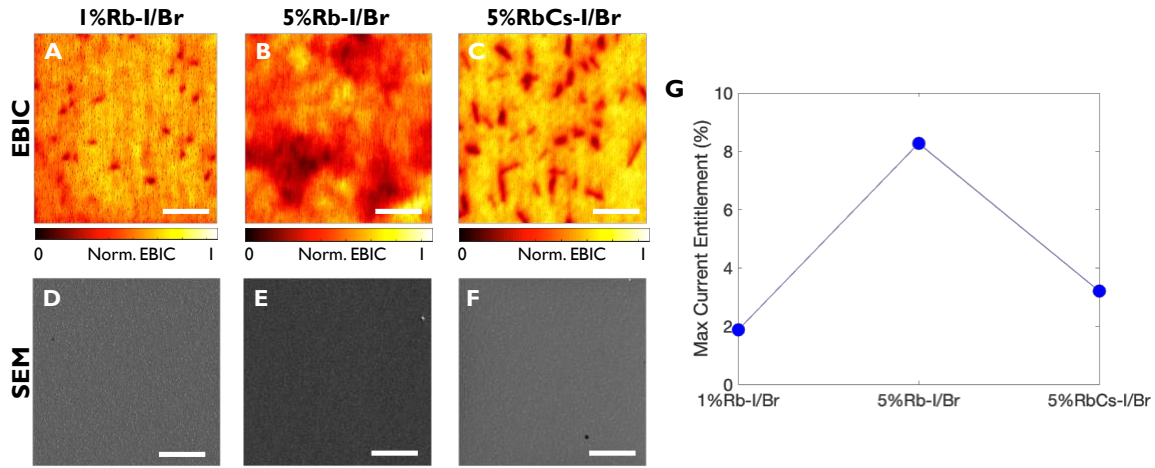

**Fig. S13. The XRD patterns of the expected perovskite and relevant secondary phases.** The RbPbI$_3$ and Rb containing perovskite phases are obtained from report published by Saliba *et al.*[3] The RbPb$_2$Br$_5$ phase is plotted with pattern number: 96-101-0212. PbI$_2$ and PbBr$_2$ has pattern number, 96-101-0063 and 96-101-0620.

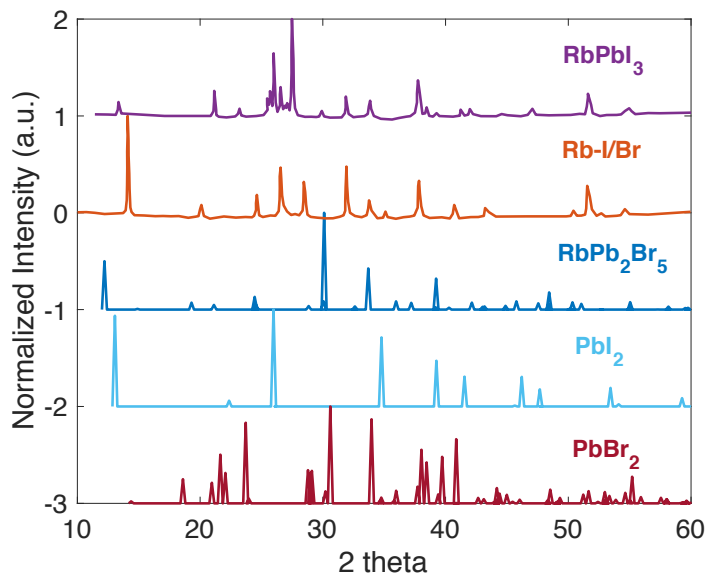